\documentclass[pra,aps,twocolumn,showpacs]{revtex4}
\usepackage{graphicx}
\usepackage{dcolumn}

\begin{document}
\def\be{\begin{equation}}
\def\ee{\end{equation}}
\def\bearr{\begin{eqnarray}}
\def\eearr{\end{eqnarray}}
\def\la{\langle}
\def\ra{\rangle}
\def\l{\left}
\def\r{\right}

\title{Quantum information entropies of ultra-cold atomic gases in a harmonic trap}
\author{Tutul Biswas and Tarun Kanti Ghosh}
\affiliation{Department of Physics, Indian Institute of Technology-Kanpur, 
Kanpur 208 016, India }
\date{\today}

\begin{abstract}
The position and momentum space information entropies of weakly 
interacting trapped atomic Bose-Einstein condensates and 
spin-polarized trapped atomic Fermi gases at absolute zero temperature
are evaluated. 
We find that sum of the position and momentum space information entropies 
of these quantum systems contain $N$ atoms confined in a $D(\le 3)$-dimensional 
harmonic trap has a universal form as 
$ S_t^{(D)} = N(a D  - b \ln N) $, 
where $a \simeq 2.332$ and $b = 2$ for interacting bosonic systems 
and $a \simeq 1.982$ and $b = 1$ for ideal fermionic systems. 
These results obey the entropic uncertainty relation given by 
the Beckner, Bialynicki-Birula and Myceilski.

\end{abstract}
\pacs{67.85.-d, 89.70.Cf, 03.65.Ta}
\maketitle

Keywords: ultra-cold atomic gases, information entropy, foundations of quantum mechanics\\

Corresponding Author: Tarun Kanti Ghosh\\
Permanent Address: Department of Physics, Indian Institute of Technology-Kanpur, Kanpur-208 016 \\
E-mail: tkghosh@iitk.ac.in\\
Phone: +91-512-259 7276\\  
Fax: +91-512-259 0914

\section{Introduction}
An information entropy, also called Shannon information entropy, 
is defined as $S_r = - \int n({\bf r}) \ln n({\bf r}) d{\bf r} $
in position space and $S_k = - \int n({\bf k}) \ln n({\bf k}) d{\bf k}$ 
in momentum space.
Here, $ n({\bf r}) = | \psi({\bf r}) |^2 $ is the density in position 
space and $ n({\bf k}) = | \psi({\bf k}) |^2 $ is the density 
in momentum space. 
These two entropies play an important role 
to provide an alternative and stronger version of the 
Heisenberg uncertainty relations.
The entropy-based uncertainty relations were first conjectured
by Hirschman \cite{hirsch} and proved by Bialynicki-Birula and 
Myceilski and independently by Beckner \cite{bbm}.
It was proved that sum of these two entropies of a single
particle obeys the entropic uncertainty relation: 
$ S_t = S_r + S_k \geq D (1 + \ln \pi) $, where $ D $ is the 
dimension of the given system. The above equation is known
as Beckner, Bialynicki-Birula and Mycielski (BBM) inequality in
the literature.
For $ N $ number of particles in the system, the above inequality 
becomes \cite{gadre0}
$ S_t \geq N [D (1 + \ln \pi) - 2  \ln N] $.
Here, the total entropy $S_t$ is negative for large $N$ because of 
the fact that the density $n({\bf r})$ is normalized to $N$. If the 
density is normalized to unity, then the total entropy becomes 
$S_t \geq [D (1 + \ln \pi) + 2  \ln N]$.
Although $S_r$ and $S_k$ are individually unbounded but their
sum is bounded from below. Therefore, the total information entropy 
(TIE) can not be decreased beyond a limit as given by the inequality.
The above inequality captures the physical fact that a highly localized 
probability distribution $ n({\bf r}) $ in real space is associated with 
a delocalized probability distribution $n({\bf k})$ in momentum space and 
vice versa.
The TIE is saturated to the lower bound of the BBM inequality when the 
reciprocity between position and momentum spaces does not hold.
Quantum harmonic oscillator (QHO) is one such example where the density
in real and momentum spaces has the same Gaussian shape. In fact, 
it has been shown in Refs. \cite{majernik,dehesa_hatom} that the TIE of the ground
state of a QHO is saturated to the lower bound of the BBM inequality:
$ S_t = D(1 + \ln \pi)$. 
It is also interesting to mention that the TIE is invariant under 
the scale transformations \cite{gadre0}. The net information content in 
$S_t$ does not modify even if we alter the wave function by uniform 
stretching or compression.

The position and momentum space entropies have been calculated analytically 
for few quantum mechanical systems. 
The position and momentum space entropies of a simple harmonic oscillator in
the ground state were exactly calculated \cite{majernik}. The information 
entropies have been calculated for $D$ dimensional harmonic oscillator and 
Coulomb problems \cite{dehesa_hatom}.
In Ref. \cite{well}, the entropic uncertainty relations of a particle in an 
infinite potential well have been analyzed. Gadre \cite{gadre} computed 
$ S_t $ for three-dimensional (3D) uniform neutral many-electron atomic systems 
within the Thomas-Fermi (TF) approximation. The TIE of 3D uniform spin-polarized 
ideal Fermi systems has been studied \cite{massen_fermi} analytically. 
In all these cases, the TIE is obeys the BBM inequality. Information entropies in 
various fields, e.g. mathematical physics, nuclear physics, chemical physics, 
information theory and other areas of physics have been extensively studied in 
recent past \cite{massen_nuclei,abe,majernik2,panigrahi,santhanam,bala,agarwal}.

There is a continuous interest on various aspects of the ultra-cold alkali 
atomic gases confined in a harmonic trap. 
It would be interesting to test whether the TF density profile, 
which is being used to describe the ultra-cold alkali atomic gases successfully, 
obey the BBM inequality or not.
To the best of our knowledge, there has not been any study on the information 
entropies of these quantum systems. The present article is devoted to the study 
of the information entropies of weakly interacting trapped atomic Bose-Einstein 
condensates (BEC) and spin-polarized trapped atomic Fermi gases at absolute zero 
temperature. We find that sum of the position and momentum space entropies has a 
universal form as $ S_t^{(D)} = N(a D  - b \ln N) $ for $D$-dimensional bosonic 
systems as well as for the fermionic systems. 
For interacting bosonic systems, we find that $ a \simeq 2.332 $ and $ b = 2 $ 
and for spin-polarized fermionic systems, $ a \simeq 1.982 $ and $ b = 1 $.  
Our results are consistent with the entropic uncertainty relation given by BBM 
\cite{bbm}.

This paper is organized as follows. In Section II, we evaluate the position and 
momentum space information entropies for weakly interacting trapped atomic BEC 
in one, two and three dimensions. In Section III, we calculate the position and 
momentum space entropies for spin-polarized fermions in a harmonic trap in one, 
two and three dimensions. We present a summary of our work in Section IV.

\section{Information Entropies of Atomic BECs}

We first consider a Bose-Einstein condensate contains $N$ number of alkali atoms 
confined in a harmonic trap $V({\bf r})$ at absolute zero temperature 
\cite{pethick_book}. This atomic system is described by the well known 
Gross-Pitaevskii equation:
\be
- \frac{\hbar^2}{2m} \nabla^2 \psi({\bf r}) 
+  V({\bf r}) \psi({\bf r}) + g |\psi({\bf r})|^2 \psi({\bf r}) 
= \mu \psi({\bf r}),
\ee
where $m$ is the mass of the alkali atom, $\mu $ is the chemical potential 
determined through the normalization condition $\int n({\bf r}) d {\bf r}=N$ 
and $ g $ is the interaction strength characterized by the two-body 
scattering length $a_s$. Also, 
$ \psi({\bf r}) = \sqrt{n({\bf r})} e^{i \phi({\bf r})} $
is the condensate wave function written in terms of the density $n({\bf r})$ 
and the phase $ \phi({\bf r})$. For typical experimental parameters, the kinetic 
energy is very very small compared to the two-body interaction energy and therefore 
the kinetic energy can be neglected safely. This approximation is known as
the TF approximation \cite{rmp}. Within the TF approximation, the density profile 
of the interacting atomic BECs in a harmonic trap
$ V (r) =  (m/2) (\omega_{\rho}^2 \rho^2 + \omega_z^2 z^2)$ is 
given by $ n({\bf r}) = (\mu - V(r))/g $, where $ \omega_{\rho} $ and 
$ \omega_z $ are the radial and axial oscillator frequencies. 
The low-dimensional BECs can be obtained by changing the trapping frequencies. 
The low dimensional BECs have also been observed experimentally 
\cite{gorlitz} and studied theoretically \cite{ketterle,ho}.

The density profile in momentum space can be obtained from the Fourier 
transformation of $ \sqrt{n({\bf r)}} $ as given by \cite{baym}
\be
n({\bf k}) = | \psi ({\bf k}) |^2 = \frac{1}{(2\pi)^D}
\left | \int \sqrt{n({\bf r})} e^{i {\bf k} \cdot 
{\bf r}} d^D {\bf r} \right |^2. 
\ee 

One can calculate the total information entropy by using the probability 
distribution in the real space as well as in the momentum space. 
Now we consider one, two and three dimensions separately and calculate the 
position and momentum space entropies for bosonic and fermionic systems.

{\it 3D Isotropic BEC}:
Let us first consider a 3D BEC confined in an isotropic harmonic 
potential ($ \omega_{\rho} = \omega_z = \omega_0 $) at very low temperature. 
The two-body interaction strength is given by 
$ g = 4 \pi a_s \hbar^2/m $ 
and the chemical potential is related to the radial size of the condensate 
$R_0 $ as $ \mu = m \omega_0^2 R_0^2/2 = 
(\hbar \omega_0/2)(15 Na_s/a_0)^{2/5}$. Here, $ a_0 = 
\sqrt{\hbar/(m\omega_0)} $ is the harmonic oscillator length.  
The TF density profile in the real space is given by
$n(r) = (1/a_0^3) (15 N/8 \pi)^{2/5} (a_0/a_s)^{3/5} 
(1- \tilde r^2)$,
where $ \tilde r = r/R_0 $ is the dimensionless variable.
Using the definition of the information entropy in 
the position space, we get
\be
S_r^{(3)} =  N \left (0.680 - \ln \frac{15 N}{8 \pi R_0^3} \right). 
\ee 
The 3D density profile in the momentum 
space is given by \cite{baym}
$ n(k) =  a_0^3 (15N/16) (15Na_s/a_0)^{3/5} 
(J_2(\tilde k)/\tilde k^2)^2 $
where $ J_n(y) $ is the $n$-th order Bessel function and 
$ \tilde k = k R_0 $.
 
Using the definition of the momentum space entropy, 
after lengthy but straightforward calculation, we obtain
\be
S_k^{(3)} \simeq  N \left (5.829 - \ln \frac{15 N R_0^3}{16} \right).
\ee
The detail calculations of $ S_r^{(3)} $ and $S_k^{(3)} $ are given
in the Appendix A.
The sum of these two entropies is given by
\be
S_t^{(3)} = S_r^{(3)} + S_k^{(3)} \simeq N(7.090 - 2 \ln N).
\ee

{\it Quasi-2D BEC}:
Now we consider a quasi-2D BEC of atomic gases. 
A quasi-2D BEC is obtained when the harmonic confinement along a 
particular direction is strong compared to the other directions 
$i. e.$ $ \omega_z \gg \omega_{\rho} $. 
In this low dimensional system, the effective interaction strength
becomes $ g_2 = 2 \sqrt{2 \pi} \hbar \omega_z a_z a_s $
and the chemical potential is 
$ \mu_2 =  m \omega_{\rho}^2 R_{\rho}^2/2 = 
\hbar \omega_{\rho} \sqrt{2\sqrt{2\pi}Na_s/a_z}$ \cite{ho,tkg0}. Here,
$a_z = \sqrt{\hbar/m\omega_z}$ is the oscillator length
and $ R_{\rho} $ is the transverse size of the bosonic system.
The TF density profile in quasi-2D bosonic system is given by
$ n(\rho) = (1/a_{\rho}^2) 
\sqrt{N a_z/((2\pi)^{3/2} a_s)} 
(1- \tilde \rho^2 )$,
where $ \tilde \rho = \rho/R_{\rho} $.
The density profile in the momentum space is calculated and it is given by
$ n(k_{\rho}) = a_{\rho}^2 
\sqrt{\sqrt{2^7/\pi} N^3a_s/a_z} 
\l[J_{3/2}(\tilde k_{\rho})/\tilde k_{\rho}^{3/2}\r]^2 $,  
where $ \tilde k_{\rho} = k_{\rho} R_{\rho} $.

Similar to the calculation of 3$D$ system, we calculate the position 
space and the momentum space entropies and 
these are given by, respectively,
\be
S_{\rho}^{(2)} = N \left(\frac{1}{2} - \ln \frac{2N }{\pi R_{\rho}^2} 
\right),
\ee 
and
\be
S_{k_{\rho}}^{(2)} \simeq N\left( 3.750 - \ln N R_{\rho}^2. 
\right)
\ee
Therefore, we obtain the TIE is
\be
S_t^{(2)} \simeq N(4.701 - 2 \ln N).
\ee

{\it Quasi-1D BEC}:
Here, we consider quasi-1D BEC which is 
achieved when $ \omega_{\rho} \gg \omega_z $.
In the quasi-1D system, the effective interaction strength
becomes $ g_1 = 2 \hbar \omega_{\rho} a_s $
and the chemical potential is
$ \mu_1 = m \omega_{z}^2 R_{z}^2/2 =
(\hbar \omega_{z}/2) (3 Na_s/a_{\rho})^{2/3}$ \cite{ho,tkg}. 
Here, $ R_z $ is the axial size of the bosonic system.
The TF density profile is given by
$ n(z) = (1/a_z) [9N^2a_{\rho}/(64a_s)]^{1/3} 
(1 - \tilde z^2 ) $,
where $ \tilde z = z/R_z $.
We calculate the corresponding density profile in the momentum space, 
which is given by
$ n(k_z) = a_z(3N \pi^3a_s/(512 a_{\rho}))^{1/3}  
(J_{1}(\tilde k_z)/\tilde k_z)^2 $,
where $ \tilde k_z = k_z R_z $.

Similar to the calculations for $3D$ and $2D$ systems, the position 
space and momentum space entropies are given by, respectively,
\be
S_z^{(1)} = N \left(0.568 - \ln \frac{N}{R_z} \right),
\ee 
and
\be
S_k^{(1)} \simeq N\left(1.764 - \ln N R_z \right).
\ee

The TIE for quasi-1$D$ BEC system is
\be
S_t^{(1)} \simeq N(2.332 - 2 \ln N).
\ee

{\it Discussion}:
It is interesting to note that the individual entropies, $S_r $ 
and $ S_k $, in three different dimensions explicitly depend on the 
harmonic confinement and the two-body interaction strength. 
However, the total entropies do not explicitly depend on the 
harmonic confinement and the interaction strength.

The total information entropies of weakly interacting atomic BECs 
in all the three dimensions can be written in a compact form as
\be \label{high}
S_t^{(D)} \simeq N (2.332 D - 2 \ln N).
\ee
Our result obeys entropic uncertainty relation given by BBM 
\cite{bbm} and Gadre \cite{gadre0}.

\section{Information Entropies of Atomic Fermions}
The fermionic alkali atoms have been cooled to temperatures far
below the Fermi temperature and therefore the systems reach in the
degenerate regime \cite{fermi_exp}. 
The review of the degenerate fermionic gases can be found in 
the Refs. \cite{pethick_book,fermi_rmp}.
Here, we consider $N$ spin-polarized fermions confined in a harmonic 
potential $V(r)$ at $ T = 0 $ temperature. 
Due to the inhomogeneity of the system, the local Fermi wave vector 
$ k_F({\bf r})$ is given by \cite{butts}
\be
\frac{\hbar^2 k_F(r)^2}{2m} + V(r) = E_F.
\ee
The semi-classical momentum distribution is given by \cite{butts}
\be
n({\bf k}) = \frac{1}{(2\pi)^D} \int d^D{\bf r} 
\Theta \left(k_F(r) - | {\bf k}| \right),
\ee 
where $ \Theta(x) $ is the unit step function.

{\it Fermions trapped in a 3D  isotropic harmonic potential}:
We first consider $N$ number of spin-polarized fermions confined by the
isotropic harmonic potential $ V(r) =  m \omega_0^2 r^2/2 $. 
For this system, the Fermi energy is 
$ E_F =  m\omega_0^2 R_F^2/2 = (6N)^{1/3} \hbar \omega_0 $, where
$ R_F $ is the Fermi radius.
The semi-classical density profile is \cite{butts}
$ n(r) = (1/a_0^3) \sqrt{4N/(3 \pi^4)} 
(1 - r^2/R_F^2)^{3/2} $.
The density profile in the momentum space is  \cite{butts} 
$ n(k)  = a_0^3 \sqrt{4N/(3 \pi^4)} 
(1 - k^2/K_F^2 )^{3/2} $,
where $ K_F = k_F(r=0) $ is the characteristic Fermi wave vector which 
is determined by the momentum of a free particle of energy $E_F$ as 
$\hbar K_F = \sqrt{2m E_F}$. The detail calculation of $ n(k) $ is given
in the Appendix B.

We calculate the position and momentum space entropies, which are given, 
respectively, by
\be
S_r^{(3)} = N \left[\frac{1}{4}(6 \ln4 - 5) - 
\ln \frac{8N}{\pi^2 R_F^3} \right],
\ee
and
\be
S_k^{(3)} = N \left[\frac{1}{4}(6 \ln4 - 5) -
\ln \frac{R_F^3}{6\pi^2 } \right],
\ee
The total information entropy in a 3D fermionic system is
\be
S_t^{(3)} = N ( 5.950 - \ln N).
\ee

{\it Fermions trapped in a 2D harmonic potential}:
For spin-polarized fermions in a 2D harmonic trap, the Fermi 
energy is $ E_F =   m \omega_{\rho}^2 R_{F_{\rho}}^2/2 = 
\sqrt{2N} \hbar \omega_{\rho} $ \cite{tkg0}.
The semi-classical density profile is
$ n(\rho) = (1/a_{\rho}^2) \sqrt{N/(2 \pi^2)} 
(1 - \rho^2/R_{F_{\rho}^2}) $.
We evaluate the corresponding density profile in the momentum space 
which is given by 
$ n(k_{\rho})  = a_{\rho}^2 \sqrt{N/(2\pi^2)} 
(1- k_{\rho}^2/K_{F_{\rho}}^2)$.
Similar to the above calculation for 3$D$ fermionic system,
the position space and the momentum space entropies are given, 
respectively, by
\be
S_{\rho}^{(2)} = N \left[\frac{1}{2} - 
\ln \frac{2N }{\pi R_{F_{\rho}}^2} \right],
\ee
and
\be
S_{k}^{(2)} = N \left[\frac{1}{2} - 
\ln \frac{R_{F_{\rho}}^2}{4 \pi } \right].
\ee
Therefore, the total information entropy in a 2D fermionic system is
\be
S_t^{(2)} = N( 3.982 - \ln N).
\ee

{\it Fermions trapped in a 1D harmonic potential:}
For 1D fermionic system, the Fermi energy is 
$ E_F =  m \omega_z^2 R_{F_{z}}^2/2 =  N \hbar \omega_z $.
The semi-classical density profile is
$ n(z) = (1/a_z) \sqrt{2N/\pi^2} 
(1 - z^2/R_{F_{z}}^2)^{1/2} $.
The corresponding density profile in the momentum space is
evaluated and it is given by 
$ n(k_z)  = a_z \sqrt{2N/\pi^2} 
(1- k_z^2/K_{F_z}^2 )^{1/2} $.

Similar to the above calculations for 3$D$ and 2$D$ fermionic systems, 
the position space and the momentum space entropies are given, 
respectively, by
\be
S_z^{(1)} = N \left[\frac{1}{2}(\ln 4 -1) - \ln \frac{2N}{\pi R_{F_z}} 
\right],
\ee
and
\be
S_{k_{z}}^{(1)} = N\left[ \frac{1}{2}(\ln 4 -1) - \ln \frac{R_{F_z}}{ \pi} 
\right].
\ee
Therefore, the TIE in a 1D fermion system is given by
\be
S_t^{(1)} = N(1.982 - \ln N).
\ee

{\it Discussions}:
Similar to the bosonic systems, the individual entropies, 
$S_r $ and $S_k $, in the three different dimensions depend 
on the harmonic confinement but the TIEs do not.
The total information entropy of $ N $ fermions in a $ D$-dimensional 
harmonic trap can be written in a compact form as 
\be
S_t^{(D)} \simeq N (1.982 D - \ln N).
\ee
The TIE of harmonically trapped spin-polarized fermionic atoms is larger 
than that of a harmonically trapped ideal bosonic atoms. The enhancement 
of the TIE of fermionic atoms compared to the bosonic atoms is due to the 
effect of the Pauli exclusion principle.

We have also calculated the TIE of Fermi systems confined in a harmonic 
trap exactly by using the harmonic oscillator wave functions in one, two 
and three dimensions. The exact results match very well with the semi-classical 
results for large number of atoms as expected.

The TIE of a 3D uniform non-interacting Fermi system has been studied 
\cite{massen_fermi}. We find that the TIE of a 3D non-interacting Fermi 
system confined in a harmonic trap is larger than that of a uniform 
non-interacting Fermi system. The increase in the information entropy 
is due to the effect of the harmonic confinement.

\section{Summary}
We have calculated the position and momentum space information 
entropies of weakly interacting alkali atomic BECs confined by 
quasi-1$D$, quasi-2$D$ and 3D harmonic trap. We have shown that 
the TIE of BEC in different dimensions does not explicitly depend 
on the two-body interaction strength and the harmonic confinement. 
The two-body interaction and the non-uniformity of the systems 
increase the TIE. 
We have also calculated the TIE of spin-polarized fermions confined 
by harmonic trap at zero temperature. Similar to the Bose systems, 
the TIE of fermions does not explicitly depend on the harmonic 
confinement. The effect of the Pauli exclusion principle is also 
reflected in the TIE of fermionic systems.

We found a similar universal form of the TIE as found by  
BBM and Gadre \cite{bbm,gadre0}. The universal form is 
$ S_t^{(D)} = N( a D  - b \ln N) $ for 
the bosonic system as well as fermionic system in a $D$-dimensional 
harmonic trap at $ T = 0 $. We found that $ a \simeq 2.332 $ and 
$ b = 2 $ for interacting bosons and for ideal fermionic system, 
$ a \simeq 1.982 $ and $ b = 1 $. Therefore, 
the total information entropies of the ultra-cold atomic gases described
by the TF density profiles obey the BBM inequality \cite{bbm}.

\begin{appendix}

\section{}
Here, we provide detail calculations of $S_r^{(3)}$ and $S_k^{(3)}$ for 3D 
bosonic system. The 
calculation of information entropies in lower dimensions and of fermionic systems 
will follow the same method.
The position space entropy is given by
\be \label{sr3}
S_r^{(3)} = - \int d^3r n({\bf r}) \ln n({\bf r}),
\ee
where $n({\bf r}) = (\mu - V(r))/g = (\mu/g)(1 - \tilde r^2)$ is the 3D TF density profile.
Substituting the TF density profile into the Eq. (\ref{sr3}), we get
\bearr
S_r^{(3)} & = & - \frac{4 \pi \mu R_0^3 }{g} \int_{\tilde r =0}^{1} 
d \tilde r \tilde r^2 (1 - \tilde r^2)
[\ln \frac{\mu}{g} + \ln(1-\tilde r^2)] \nonumber \\
& = & 
- \frac{4 \pi \mu R_0^3 }{g}  \ln \frac{\mu}{g} 
\int_0^1 d \tilde r \tilde r^2 (1 - \tilde r^2) \nonumber \\
& - & \frac{4 \pi \mu R_0^3 }{g}  
\int_0^1 d \tilde r \tilde r^2 (1 - \tilde r^2) \ln (1-\tilde r^2)  \nonumber \\
& = & 
- \frac{4 \pi \mu R_0^3 }{g} \l[ \frac{2}{15} \ln \frac{\mu}{g} +
\frac{2}{225}( -31 + 30 \ln2)\r] \nonumber \\
& = &
- \frac{8 \pi \mu R_0^3 }{15 g} \l[ \ln \frac{\mu}{g} - 0.680\r].
\eearr
Using the expressions for $g$, $\mu$ and $R_0$ of a 3D bosonic system given in 
the Section II, one can easily show that $ \frac{\mu R_0^3}{g} = \frac{15N}{8 \pi} $.
Hence,
\be
S_r^{(3)} = N \l (0.680 - \ln \frac{15 N }{8 \pi R_0^3} \r). 
\ee 

Using the expression for the momentum distribution $n({\bf k})$ \cite{baym}, 
the momentum space entropy in 3D is calculated as given below.
\bearr
S_k^{(3)} & = & - \int d^3k n(k) \ln n(k) \nonumber \\
& = & -\frac{15 \pi N}{4}  \ln \frac{15NR_0^3}{16} \int_0^{\infty} d\tilde k \tilde k^2 
\l|\frac{J_2(\tilde k)}{\tilde k^2} \r|^2 \nonumber \\
& -  & \frac{15 \pi N}{4}  \int_0^{\infty} d\tilde k \tilde k^2
\l|\frac{J_2(\tilde k)}{\tilde k^2}\r|^2 \ln \l|\frac{J_2(\tilde k)}{\tilde k^2}\r|^2.
\eearr
The standard result \cite{wolf} of the first integral on the right hand side of 
the above equation is
\be
\int_0^{\infty} d\tilde k \tilde k^2 \l|\frac{J_2(\tilde k)}{\tilde k^2} \r|^2 = \frac{4}{15 \pi}.
\ee
We calculate the second integral numerically. Note that 
$ J_2(x)/x^2 $ does not vanish as $x$ approaches to zero since
$ J_2(x) = x^2/8 - x^4/96 + ...$. Therefore, 
$ \ln |J_2(x)/x^2|^2 $ is finite at $x=0$.
The numerical result of the second integral is
\be
\int_0^{\infty} d\tilde k \tilde k^2
\l|\frac{J_2(\tilde k)}{\tilde k^2} \r|^2 \ln \l|\frac{J_2(\tilde k)}{\tilde k^2} \r|^2 \simeq - 0.494.
\ee
Therefore,
\be
S_k^{(3)} \simeq  N \left (5.829 - \ln \frac{15 N R_0^3}{16} \right).
\ee

\section{}

The semiclassical momentum space distribution for a harmonically trapped 
Fermi gas in 3D at $T=0$  \cite{butts} is
\bearr
n(k) & = & \frac{1}{(2\pi)^3} \int d^3r \Theta( k_F(r) - |{\bf k}|) \nonumber \\
& = & \frac{1}{2\pi^2} \int_0^{R_F} dr r^2  \Theta (k_F(r) - |{\bf k}|).
\eearr
Changing variable from $r$ to $k_F(r)$.
At $r = 0$, $k_F(r=0) = K_F = \sqrt{2m E_F/\hbar^2} $ and 
at $ r = R_F$, $k_F(r = R_F) = 0 $.
Also,
\be
r^2 dr = - a_0^6 
\sqrt{K_F - k_F^2}  k_F dk_F.
\ee

Therefore,
\bearr
n(k) & = & - \frac{a_0^6}{2\pi^2}
\int_{K_F}^0 \Theta (k_F(r) - |{\bf k}|) \sqrt{K_F - k_F^2}  k_F dk_F \nonumber \\
& = & -  \frac{a_0^6}{2\pi^2}
\int_{K_F}^{k} \Theta( k_F(r) - |{\bf k}|) \sqrt{K_F - k_F^2} k_F dk_F \nonumber \\
& - & \frac{a_0^6}{2\pi^2} \int_k^0  \Theta( k_F(r) - |{\bf k}|) \sqrt{K_F - k_F^2} 
k_F dk_F \nonumber \\
&= & -\frac{a_0^6}{2\pi^2} \int_{K_F}^k \sqrt{K_F - k_F^2} k_F dk_F \nonumber \\
& = & a_0^3 \sqrt{\frac{4N}{3 \pi^4}} 
\l(1 - \frac{k^2}{K_F^2} \r)^{3/2}.
\eearr
Here, $ k = |{\bf k}| $ is the magnitude of the wavevector ${\bf k} $.

\end{appendix}

\end{document}